\documentclass[twocolumn,superscriptaddress,showpacs,prl]{revtex4}
\usepackage{mathrsfs}
\usepackage{amssymb, amsbsy, amsmath, latexsym, dsfont, array, layout, graphicx,mathrsfs,color}

\newcommand{\one}{\mathds{1}}
\newcommand{\ket}[1]{\left|{#1}\right\rangle}
\newcommand{\bra}[1]{\left\langle{#1}\right|}

\begin{document}

\title{Experimental realization of a single qubit SIC POVM on via a one-dimensional photonic quantum walk}
\author{Zhihao Bian}
\affiliation{Department of Physics, Southeast University, Nanjing
211189, China}
\author{Jian Li}
\affiliation{Department of Physics, Southeast University, Nanjing
211189, China}
\author{Hao Qin}
\affiliation{Department of Physics, Southeast University, Nanjing
211189, China}
\author{Xiang Zhan}
\affiliation{Department of Physics, Southeast University, Nanjing
211189, China}
\author{Peng Xue\footnote{gnep.eux@gmail.com}
} \affiliation{Department of Physics, Southeast University, Nanjing
211189, China} \affiliation{State Key Laboratory of Precision
Spectroscopy, East China Normal University, Shanghai 200062, China}
\affiliation{Beijing Institute of Mechanical and Electrical Space}

\begin{abstract}
Symmetric informationally complete positive operator-valued measurement (SIC-POVM) is one important class of quantum measurement which is crucial for various quantum information processing tasks. SIC-POVMs have the advantage of providing an unbiased estimator for quantum states with the minimal number of outcomes needed for full tomography. We present an experimental approach on a photonic quantum walk which can be used to implement SIC-POVMs on a single-qbubit. The projection measurements of single-photons' positions correspond to elements of SIC-POVM on the polarization of single-photons.

\end{abstract}
\pacs{03.65.Ta, 03.67.Ac, 03.67.Lx, 42.50.Dv, 42.50.Xa}

\maketitle

Symmetric informationally complete positive operator-valued measurement (SIC-POVM) is a generalized measurement on a $d$-dimensional quantum state, which consists of $d^2$ subnormalized projection operators with equal pairwise fidelity~\cite{RBS+04,SG10}. In practical quantum state tomography, SIC-POVM provides an elegant description of quantum states and is regarded as the most efficient and optimal technique for characterizing quantum states. A SIC-POVM can always be implemented with a suitable coupling between the system and an auxiliary system and by performing a projective measurement on the joint system. Experimentally SIC-POVM has been realized on photonic~\cite{LSLK06,DKLL08,MTS+11} and NMR qubits~\cite{DSPD06} which are allowed to couple to ancillary systems and to interact and measure jointly. 

Recently Kurzy\'{n}ski and W\'{o}jcik propose an idea~\cite{KW13} that one-dimensional discrete-time quantum walk can approximate the von Neumann model that is extended in time and for which the internal evolution of the measured coin affects the evolution of the walker of the measuring device. Since the idea of quantum walks
originates from the von Neumann model of measurement~\cite{ADZ93}, in which the change of the position of the
walker depends on the state of the coin that needs to be measured, one can naturally extend the von Neumann measurement model to an arbitrary SIC-POVM without the requirement of extension of the Hilbert space.

A single step of an initially localized discrete-time quantum walk~\cite{K03,K06,COR+13,SCP+11,GAS+13,BFL+10,ZKG+10} can be considered as a projective von Neumann measurement of the coin qubit for the walker is found in the position $x=i$ corresponding to the coin being in the certain state, i.e., the projection measurement of the walker's position corresponds to SIC-POVM on the coin state.

A qubit SIC-POVM corresponds to a set of projection operators associated with the vertices of a regular tetrahedron circumscribed by the Bloch sphere. The choice for the SIC-POVM is not unique since one can always rotate the Bloch sphere but we may choose the projectors corresponding to the set of vectors $\{\xi_i\}^4_{i=1}\in \mathcal{C}^2$, which satisfy $|\langle \xi_i,\xi_j\rangle|=1/\sqrt{3}$ for $i\neq j$. A POVM generated by the above vectors $\{1/2\ket{\xi_i}\bra{\xi_i}\}_{i=1}^4$ is so called a SIC-POVM, and hence $\sum_i^4 1/2\ket{\xi_i}\bra{\xi_i}=\one$ is satisfied. In $\mathcal{C}^2$,
\begin{align}
&\xi_1=\left(
        \begin{array}{c}
          1 \\
          0 \\
        \end{array}
      \right),
\xi_2=\frac{1}{\sqrt{3}}\left(
        \begin{array}{c}
          1 \\
          \sqrt{2} \\
        \end{array}
      \right),\nonumber\\
&\xi_3=\frac{1}{\sqrt{3}}\left(
        \begin{array}{c}
          1 \\
          \lambda\sqrt{2} \\
        \end{array}
      \right),
      \xi_4=\frac{1}{\sqrt{3}}\left(
        \begin{array}{c}
          1 \\
          \lambda^*\sqrt{2} \\
        \end{array}
      \right)
\end{align}
is a SIC-POVM, where $\lambda=e^{i2\pi/3}=(-1+\sqrt{3}i)/2$.

This single qubit SIC-POVM can be realized by using a photonic quantum walk, where the information is encoded in the horizontal and vertical polarization of photons representing the quantum coin. A SIC-POVM on the coin state can be implemented by controlling the internal dynamics of the coin state. In our experiment the the internal dynamics of the coin is controlled by site-dependent coin flipping.

Firstly we find four states which are orthogonal to the states in Eq.~(1) respectively and prepare the coin states in the four states initially. Thus the algorithm starts with four coin states as following:
\begin{align}
&\ket{\psi_1}=\ket{V},\nonumber\\
&\ket{\psi_2}=\frac{1}{\sqrt{3}}(\sqrt{2}\ket{H}-\ket{V}),\nonumber\\
&\ket{\psi_3}=\frac{1}{\sqrt{3}}(\sqrt{2}\ket{H}-\lambda\ket{V}),\nonumber\\
&\ket{\psi_4}=\frac{1}{\sqrt{3}}(\sqrt{2}\ket{H}-\lambda^*\ket{V}),
\end{align}
where $\ket{H}=(1,0)^T$ and $\ket{V}=(0,1)^T$ are horizontal and vertical polarization of photons respectively, and $\langle\psi_i|\xi_i\rangle=0$ is satisfied. 

Now let us consider an algorithm for generation of a qubit SIC-POVM via a split-step quantum walk. Here we use a photonic set-up demonstrated in \cite{XQTS14,XQT14} to implement a variation of these walks, the split-step quantum walk of single photons, with two qubits of the coin states encoded in its horizontal $\ket{H}$ and vertical $\ket{V}$ polarization states.

\begin{figure}
\includegraphics[width=8.5cm]{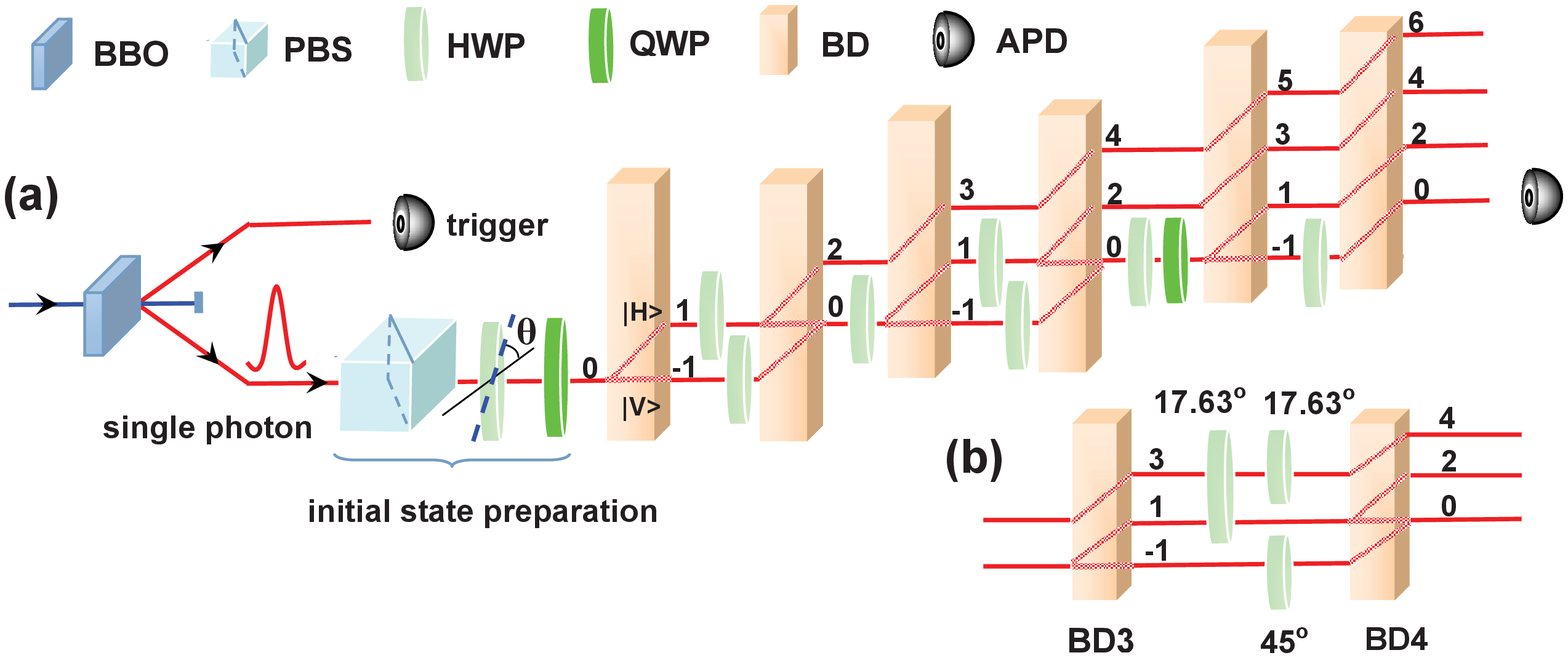}
\caption{(Color online.) (a) Experimental schematic. Detailed sketch of
the setup for realization of a qubit SIC-POVM via a split-step quantum walk.
Single photons are created via SPDC in a BBO crystal. One photon
in the pair is detected to herald the other photon, which is
injected into the optical network. Arbitrary initial coin states are
prepared by a PBS and wave plate. Site-dependent coin flipping is
realized by HWP and QWP with different setting angles placed in different
optical paths. The conditional position shift is implemented by BDs. Coincident detection of trigger and heralded photons
is done via avalanche photo-diodes to yield data for the quantum walk. (b) The detailed experimental setup for the interferometer formed by the third and fourth BDs. }\label{Fig1}
\end{figure}

\begin{table}[htbp]
\begin{tabular}{l|l|l|l|l|l|l}
  \hline
   & HWP$_0$ & QWP$_0$ & HWP$_1$ & QWP$_1$& HWP$_{-1}$ & QWP$_{-1}$\\ \hline \hline
  step 1 & $-$ & $-$ & $-22.5^o$& $-$ & $45^o$ & $-$ \\ \hline
  step 2 & $67.5^o$ & $-$ & $17.63^o$ & $-$ & $45^o$ & $-$ \\ \hline
  step 3 & $52.5^o$ & $45^o$ & $-$ & $-$ & $45^o$ & $-$ \\
  \hline
\end{tabular}
\caption{The setting parameters of the HWPs and QWPs which are used to realize site-dependent coin rotations for the three-step split-step quantum walk. The subscripts denote the optical modes where the wave plates are placed and ``$-$" means no corresponding wave plate is used.}
\end{table}

The polarization degenerate photon pairs are generated via type-I
spontaneous parametric down-conversion (SPDC) in $0.5$mm-thick
nonlinear-$\beta$-barium-borate (BBO) crystal which is pumped by a $400.8$nm CW diode laser with $90$mW of power.
For one-dimensional quantum walks, triggering on one photon prepares the other photon pairs
at wavelength $801.6$nm into a single-photon state in Eq.~(2). The photon bandwidth is $3$nm determined with the interference filter.
The total coincident counts are about $3.2\times10^4$ and the coincident
counts are collected over $60$s. The probability of creating more
than one photon pair is less than $10^{-4}$ and can be neglected.

The initial coin state in Eq.~(2) can be prepared by the half-wave plate (HWP) or quatre-wave plate (QWP) right after the polarizing beam splitter (PBS) shown in Fig.~1(a). For example, the initial coin states $\ket{\psi_1}$ and $\ket{\psi_2}$ can be prepared by a HWP with angles set to $\theta_H=45^o$ and $-17.63^o$, and the other two initial coin states $\ket{\psi_3}$ and $\ket{\psi_4}$ can be prepared by setting angle of QWP to $\theta_Q=-152.63^o$ and $117.37^o$, respectively. The single-qubit rotations realized by HWP and QWP are
\begin{align}
&R_{HWP}(\theta_H)=\begin{pmatrix}\cos2\theta_H & \sin2\theta_H\\
\sin2\theta_H & -\cos2\theta_H\end{pmatrix},\\\nonumber
&R_{QWP}(\theta_Q)=\begin{pmatrix}\cos^2\theta_Q+i \sin^2 \theta_Q & (1-i)\sin\theta_{Q}\cos\theta_Q\\
(1-i)\sin\theta_Q\cos\theta_Q & \sin^2\theta_Q+i\cos^2\theta_Q \end{pmatrix},
\end{align}
respectively, where $\theta_H$ and $\theta_Q$ are the angles between the optic axes of HWP and QWP and horizontal direction.

The quantum walk takes place on a one-dimensional lattice and starts in the position $x=0$. The walker's positions are represented by longitudinal spatial
modes. The down-converted photons are steered into the optical modes of the linear-optical network formed by
a series of birefringent calcite beam displacers (BDs) and wave
plates. One step of the split-step quantum walk consists of two sub-steps. In the $i$th step, for the first sub-step site-dependent rotation on polarization of single photon $C^i_1(0)\in$SU(2)
in the position $x=0$ is achieved by HWPs and QWPs with certain setting angles placed in the optical mode $x=0$, and identity elsewhere. The subscript $1$ denotes the index of sub-step, i.e. the first sub-step. Then a polarization-dependent position shift operation
\begin{equation}
T=\sum_{x}\ket{x+1}\bra{x}\otimes\ket{H}\bra{H}+\ket{x-1}\bra{x}\otimes\ket{V}\bra{V}
\end{equation} is applied on the photons with $\ket{H}$ ($\ket{V}$) to the right (left) by one lattice site using a BD. This is followed by the second sub-step including site-dependent rotations $C^i_2(1)$ in the position $x=1$ and $C^i_2(-1)$ in the position $x=-1$ and identity elsewhere, and finally another position shift translation $T$. The $i$th-step quantum walk is implemented by repeating applications of the step operator
\begin{align}
U_i=&T\big[|1\rangle\langle 1|\otimes C^i_2(1)+|-1\rangle\langle -1|\otimes C^i_2(-1)\\&+\sum_{x\neq \pm 1}|x\rangle\langle x|\otimes \one\big]T
\big[|0\rangle\langle 0|\otimes C^i_1(0)+\sum_{x\neq 0}|x\rangle\langle x|\otimes \one\big].\nonumber
\end{align}

In this scenario a single-qubit SIC-POVM can be realized
via a three-step split-step quantum walk. For the first step, the site-dependent coin operations are chosen as
\begin{equation}
C^1_1(0)=\one, C^1_2(1)=\frac{1}{\sqrt{2}}\left(
         \begin{array}{cc}
           1 & -1 \\
           -1 & -1 \\
         \end{array}
       \right), C^1_2(-1)=\sigma_x.
\end{equation}
The site-dependent coin operators for the second step are
\begin{align}
&C^2_1(0)=\frac{1}{\sqrt{2}}\left(
         \begin{array}{cc}
           -1 & 1 \\
           1 & 1 \\
         \end{array}
       \right),
       C^2_2(1)=\frac{1}{\sqrt{3}}\left(
         \begin{array}{cc}
           \sqrt{2} & 1 \\
           1 & -\sqrt{2} \\
         \end{array}
       \right), \nonumber\\
       &C^2_2(-1)=\sigma_x.
\end{align}
Whereas, for the third step, we have
\begin{equation}
C^3_1(0)=\frac{1}{\sqrt{2}}\left(
         \begin{array}{cc}
           e^{-i\frac{\pi}{3}} & e^{i\frac{\pi}{6}} \\
           e^{i\frac{\pi}{3}} & e^{-i\frac{\pi}{6}} \\
                    \end{array}
       \right),
       C^3_2(1)=\one,
       C^3_2(-1)=\sigma_x.
\end{equation} Here $\sigma_x=\left(
         \begin{array}{cc}
           0 & 1 \\
           1 & 0 \\
         \end{array}
       \right)$ is one of the Pauli operators. The site-dependent coin rotations can be realized by HWP and QWP with specific angles (show in Table I) placed in certain modes.

After three steps, the states of the walker+coin system evolve to
\begin{equation}
\ket{\varphi_i(3)}=U_3U_2U_1\ket{\varphi_i(0)},
\end{equation}
where $\ket{\varphi_i(0)}=\ket{0}\ket{\psi_i}$ ($i=1,2,3,4$) denotes the initial state of the walker+coin system.
\begin{align}
&\ket{\varphi_1(3)}=\frac{1}{\sqrt{3}}\left(-\ket{4}-i\ket{2}+i\ket{0}\right)\ket{H},\nonumber\\
&\ket{\varphi_2(3)}=\frac{1}{\sqrt{3}}\left(\ket{6}-e^{-i\frac{\pi}{3}}\ket{2}-e^{i\frac{\pi}{3}}\ket{0}\right)\ket{H},\nonumber\\
&\ket{\varphi_3(3)}=\frac{1}{\sqrt{3}}\left(\ket{6}-e^{-i\frac{\pi}{6}}\ket{4}-\ket{2}\right)\ket{H},\nonumber\\
&\ket{\varphi_4(4)}=\frac{1}{\sqrt{3}}\left(\ket{6}-e^{i\frac{\pi}{6}}\ket{4}-\ket{0}\right)\ket{H}.
\end{align}
Compared to the initial states of the walker+coin system, it is obvious that the final state $\ket{\varphi_1(3)}$ can only be found in the positions $x=0,2,4$ and not in $x=6$. The state $\ket{\varphi_2(3)}$ can not be found in the position $x=4$, $\ket{\varphi_3(3)}$ in $x=0$ and $\ket{\varphi_4(3)}$ in $x=2$. Thus we realize all the elements $\ket{\xi_i}\bra{\xi_i}/2$ ($i=1,2,3,4$) of a qubit SIC-POVM.

For the site-dependent coin flipping, the challenge is how to place the wave plate in the certain optical mode and not to influence the photons in the other modes. For example, in Fig.~1(b) for the second sub-step of the first step, the polarizations of photons in the modes $x=\pm1$ are rotated by a HWP with setting angle $\theta_H=17.63^o$ and $\theta_H=45^o$ respectively and the photons in those two modes interfere in the mode $x=0$ at the fourth BD. Because of the small separations between the neighboring modes, it is difficult to inset a HWP in the middle mode $x=1$ and avoid the photons in the neighboring modes passing through it. In our experiment, we place a HWP with $\theta_H=17.63^o$ in both modes $x=1$ and $x=3$ following by a HWP with same angle in the mode $x=3$ and a HWP with $45^o$ in the mode $x=-1$. Thus the photons in modes $x=\pm1$ do not suffer an optical delay and interfere with each other with a high visibility. The polarizations of photons in the mode $x=3$ are not changed after two HWPs with same angle. The photons in the mode $x=3$ do not interfere with those in the other modes, though there is optical delay between them. Hence no optical compensate is needed.

The conditional position shift is implemented by a BD with length $28$mm
and clear aperture $10$mm$\times 10$mm which is cut so that vertically polarized photons are directly transmitted
and horizontal photons move up a $2.7$mm lateral displacement into
a neighboring mode and interfere with the vertical photons in the
same mode. BDs are placed
in sequence. Based on the choices of the coin operation parameters shown in Table I, only two pairs of BDs, i.e. the first and second, and the third and fourth BDs form an interferometer, respectively. Thus these two pairs of BDs need to have their optical axes mutually aligned in order to ensure that beams split by one BD in the sequence
yield maximum interference visibility after passing through wave plates
and the next BD in the sequence. In our experiment, we attain interference visibility of $0.992$ on average. Coincident detection of trigger and heralded photons at avalanche photo-diodes
($7$ns time window) with dark count rate of $<100$s$^{-1}$ yields data for the quantum walk shown in Table II.

\begin{table}[htbp]
\begin{tabular}{l|l|l|l|l|l}
  \hline
  $ $ & $P(0)$ & $P(2)$& $P(4)$ & $P(6)$ & $d$ \\
  \hline\hline
  $\ket{\psi_1}$ & $0.3246(37) $ & $0.3277(38)$ &  $0.3327(38)$ & $0.0149(07)$ & $0.0149(33)$ \\ \hline
  $\ket{\psi_2}$ & $0.3398(38)$ & $0.3135(36)$  & $0.0345(11)$ & $0.3123(36)$ & $0.0401(32)$ \\ \hline
  $\ket{\psi_3}$ & $0.0335(10)$ & $0.3137(36)$ & $0.3432(38)$ & $0.3104(36)$ & $0.0425(32)$ \\ \hline
  $\ket{\psi_4}$ & $0.3158(36)$ & $0.0329(10)$ & $0.3419(38)$  & $0.3094(35)$ & $0.0415(32)$ \\ \hline
\end{tabular}
\caption{The measured probability distribution of the walker $P(x)$ and
1-norm distance from the theoretical predictions are shown corresponding to four different
initial coin states. Error bars indicate the statistical uncertainty.}
\end{table}

\begin{figure}
\includegraphics[width=8.5cm]{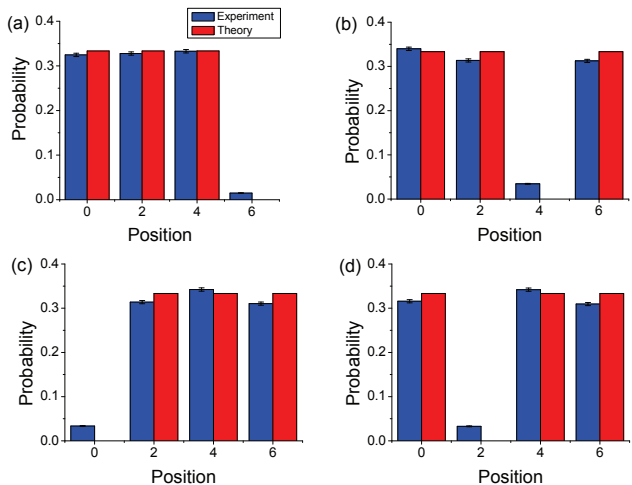}
\caption{(Color online.) Experimental data of a qubit SIC-POVM via a photonic quantum walk.
Measured probability distributions of the three-step quantum walk with the site-dependent coin rotations and four different initial
coin states $\ket{\psi_i}$ with $i=1,2,3,4$ in (a)-(d) respectively.
The blue and red bars show the experimental data and theoretical
predictions, respectively.}\label{Fig2}
\end{figure}

The realization of a qubit SIC-POVM via split-step quantum walk
in detailed are shown in Fig.~1. The measured probability distributions of quantum walk after three steps are shown in Fig.~2. The probabilities are obtained by
the normalizing coincidence counts on each mode to total for the
respective step. The measured probability distributions of three-step quantum walk for the four initial coin states agree well with the theoretical predictions. Using the experimental distribution for the initial coin state $\ket{\psi_1}$ as an example, the probability of the photons are measured in the mode $x=6$ is $0.0149\pm0.0007$ and very small compared to the probabilities of the photons being measured in the other modes, which ensures that one of the elements of a qubit SIC-POVM is realized successfully. In our experiment, the inaccuracy of angles of the wave plates and the unperfect interference visibility are the dominant source of errors. We also characterize the quality of the experimental quantum walk
by its 1-norm distance~\cite{BFL+10} from the theoretical predictions
according to
$d=\frac{1}{2}\sum_x\left|P^\text{exp}(x)-P^\text{th}(x)\right|$
shown in Table II. In our experiment the small distances ($d<0.043$)
demonstrate strong agreement between the measured distribution and
theoretic prediction after three steps.

Therefore we experimentally prove that discrete-time quantum walks are capable of
performing generalized measurements on a single qubit. We explicitly
for the first time realize a photonic three-step split-step quantum walk with
site-dependent coin flipping for a qubit SIC-POVM with
single photons undergoing an interferometer network. The SIC-POVM is confirmed by direct measurement and found to be
consistent with the ideal theoretical values at the level of the
average distance.

We use a very simple experiment to show the meaning of the idea on
implementation of a generalized measurement via a discrete-time quantum walk proposed in~\cite{KW13}. In a quantum walk, the position shifts of the walker depend on
the coin state being measured. That is, a quantum walk is based on projection
measurement which in the quantum walk scheme can be extended in the evolution
time. Furthermore as shown in the above experiment, the projection
measurement model can naturally be extended to an arbitrary
generalized measurement without the extension of the Hilbert space.
Moreover our quantum walk interferometer network can be extended to higher dimensions and the coin space can be extended to a larger Hilbert space expanded by higher-dimensional qudits. This technique can be used to realize a higher-dimensional qudit SIC-POVM which is one of the flagship measurements in quantum information processing.

\acknowledgements We would like to thank P. Kurzy\'{n}ski, C. F. Li and Y. S. Zhang for
stimulating discussions. This work has been supported by National Natural Science Foundation of China under
11174052 and 11474049, the Open Fund from the State Key Laboratory of Precision Spectroscopy of East China Normal University, the National Basic Research Development Program of China (973 Program) under Grant No. 2011CB921203 and CAST Innovation fund.


\begin{references}
\bibitem{RBS+04} J. M. Renes, R. Blume-Kohout, A. J. Scott, and C. M. Caves, J. Math. Phys. {\bf 45}, 2171 (2004).
\bibitem{SG10} A. J. Scott, and M. Grassl, J. Math. Phys. {\bf 51}, 042203 (2010).
\bibitem{LSLK06} A. Ling, K. P. Soh, A. Lamas-Linares, and C. Kurtsiefer, Phys. Rev. A {\bf74}, 022309 (2006).
\bibitem{DKLL08} T. Durt, C. Kurtsiefer, A. Lamas-Linares, and A. Ling, Phys. Rev. A {\bf 78}, 042338 (2008).
\bibitem{MTS+11} Z. E. D. Medendorp, F. A. Torres-Ruiz, L. K. Shalm,
G. N. M. Tabia, C. A. Fuchs, and A. M. Steinberg, Phys.
Rev. A {\bf83}, 051801R (2011)
\bibitem{DSPD06} J. Du, M. Sun, X. Peng, and T. Durt, Phys. Rev. A {\bf 74}, 042341 (2006).
\bibitem{KW13} P. Kurzy\'{n}ski, and A. W\'{o}jcik, Phys. Rev. Lett. {\bf 110}, 200404 (2013).
\bibitem{ADZ93} Y. Aharonov, L. Davidovich, and N. Zagury, Phys. Rev. A
{\bf48}, 1687 (1993).
\bibitem{K03} J. Kempe, Cont. Phys. {\bf44}, 307 (2003).
\bibitem{K06} V. Kendon, Math. Struct. in Comp. Sci. {\bf 17}, 1169 (2006).
\bibitem{COR+13} A. Crespi, R. Osellame, R. Ramponi, V. Giovannetti,
R. Fazio, L. Sansoni, F. De Nicola, F. Sciarrino, , and
P. Mataloni, Nat. Photon. {\bf7}, 322 (2013).
\bibitem{SCP+11} A. Schreiber, K. N. Cassemiro, V. Poto\v{c}ek, A. G\'{a}bris, I. Jex, and C. Silberhorn, Phys. Rev. Lett. {\bf106}, 180403 (2011).
\bibitem{GAS+13} M. Genske, W. Alt, A. Steffen, A. H. Werner, R. F.
Werner, D. Meschede, and A. Alberti, Phys. Rev. Lett. {\bf110}, 190601 (2013).
\bibitem{BFL+10} M. A. Broome, A. Fedrizzi, B. P. Lanyon, I. Kassal,
A. Aspuru-Guzik, and A. G. White, Phys. Rev. Lett.
{\bf104}, 153602 (2010).
\bibitem{ZKG+10} F. Z\"{a}hringer, G. Kirchmair, R. Gerritsma, E. Solano, R. Blatt, and C. F. Roos, Phys. Rev. Lett. {\bf104}, 100503 (2010).
\bibitem{XQTS14} P. Xue, H. Qin, B. Tang, and B. C. Sanders, New J. Phys. {\bf 16}, (2014).
\bibitem{XQT14} P. Xue, H. Qin, and B. Tang, Sci. Rep. {\bf4}, 4825 (2014).
\end{references}
\end{document}